\documentclass[twocolumn,aps,superscriptaddress,showpacs,preprintnumbers,amsmath,amssymb]{revtex4}
\usepackage{graphicx,color}% Include figure files
\usepackage{dcolumn}% Align table columns on decimal point
\usepackage{colordvi} % for color text
\usepackage{bm}% bold math
\begin{document}
%%%%%%%%%%%%%%%%%%%%%%%%%%%%%%%%%%%%%%%%%%%%%%%%%%%%%%%%%%%%%%%%%%%%
%\title{Measuring the Hubble parameter with gravitational-wave standard sirens}
\title{Tracing the redshift evolution of Hubble parameter with 
gravitational-wave standard sirens}
%%%%%%%%%%%%%%%%%%%%%%%%%%%%%%%%%%%%%%%%%%%%%%%%%%%%%%%%%%%%%%%%%%%%
\author{Atsushi Nishizawa}
\email{anishi@yukawa.kyoto-u.ac.jp}
\affiliation{Yukawa Institute for Theoretical Physics, Kyoto University, 
Kyoto 606-8502, Japan}
\author{Atsushi~Taruya}
\affiliation{Research Center for the Early Universe, Graduate School of 
Science, The University of Tokyo, Tokyo 113-0033, Japan}
\affiliation{Institute for the Physics and Mathematics of the Universe, 
The University of Tokyo, Kashiwa, Chiba 277-8568, Japan}
\author{Shun Saito}
\affiliation{Department of Physics, Graduate School of Science, 
The University of Tokyo, Tokyo 113-0033, Japan}
\affiliation{Department of Astronomy, 601 Campbell Hall, University of 
California Berkeley, California 94720, USA}
%%%%%%%%%%%%%%%%%%%%%%%%%%%%%%%%%%%%%%%%%%%%%%%%%%%%%%%%%%%%%%%%%%%%
\date{\today}

%%%%%%%%%%%%%%%%%%%%%%%%%%%%%%%%%%%%%%%%%%%%%%%%%%%%%%%%%%%%%%%%%%%%
\begin{abstract}
Proposed space-based gravitational-wave detectors such as BBO and DECIGO 
can detect $\sim10^6$ neutron star (NS) binaries and determine the luminosity distance to the binaries with high precision. Combining the luminosity 
distance and electromagnetically-derived redshift, one would be able to 
probe cosmological expansion out to high redshift. In this paper, we show 
that the Hubble parameter as a function of redshift can be directly measured 
with monopole and dipole components of the luminosity distance 
on the sky. As a result, the measurement accuracies of the Hubble parameter 
in each redshift bin up to $z=1$ are $3-14\,\%$, $1.5-8\,\%$, and $0.8-4\,\%$  
for the observation time $1\,{\rm{yr}}$, $3\,{\rm{yr}}$, and $10\,{\rm{yr}}$, 
respectively.
\end{abstract}

%%%%%%%%%%%%%%%%%%%%%%%%%%%%%%%%%%%%%%%%%%%%%%%%%%%%%%%%%%%%%%%%%%%%
\maketitle

%%%%%%%%%%%%%%%%%%%%%%%%%%%%%%%%%%%%%%%%%%%%%%%%%%%%%%%%%%%%%%%%%%%%
%%%%%%%%%%%%%%%%%%%%%%%%%%%%%%%%%%%%%%%%%%%%%%%%%%%%%%%%%%%%%%%%%%%%
\section{Introduction}
%%%%%%%%%%%%%%%%%%%%%%%%%%%%%%%%%%%%%%%%%%%%%%%%%%%%%%%%%%%%%%%%%%%%
%%%%%%%%%%%%%%%%%%%%%%%%%%%%%%%%%%%%%%%%%%%%%%%%%%%%%%%%%%%%%%%%%%%%

Future space-based gravitational wave detectors such as DECI-hertz 
Interferometer Gravitational-wave Observatory (DECIGO) \cite{bib18,bib19} 
and Big-Bang Observer (BBO) \cite{bib20} (also see \cite{bib6} for updated 
information) are the most sensitive to a gravitational wave (GW) in $0.1-1\,{\rm{Hz}}$ band and 
will aim at detecting a cosmological GW background generated 
during the inflationary epoch, the mergers of intermediate-mass black holes, and a large number of neutron star (NS) binaries in an inspiraling 
phase. These GW sources enable us to measure the cosmological expansion 
with unprecedented precision \cite{bib6}, to investigate the population and 
formation history of compact binary objects, and to test alternative theories of gravity \cite{bib45,bib7}. 

It has been known that the continuous GW signal from a compact-binary object 
provides a unique way to measure the luminosity 
distance to the source with high precision. Such binary sources are often 
referred to as {\it{standard siren}} (analogous to the electromagnetic 
standard candle). With the redshift information determined by an electromagnetic follow-up observation, 
the standard siren can be an accurate tracer of the cosmic 
expansion \cite{bib33}. The feasibility of the standard siren relies on the determination of the redshift of each binary. The identification of a host 
galaxy by follow-up observation are 
thus crucial, indicating that a high-angular resolution is generally 
required for GW detector. In the case of DECIGO and BBO, the detectors 
orbit the Sun with a period of one sidereal year, and constitute four 
clusters, each of which consists of three spacecrafts exchanging laser beams 
with the others. The two of the four 
clusters are located at the same position to enhance correlation sensitivity 
to a gravitational wave background, and the other two are separated on the Earth orbit in order 
to enhance the angular resolution so that we can easily identify the host 
galaxy of each NS binary via the electromagnetic follow-up observations. 
Based on this setup, Cutler 
and Holz \cite{bib6} have shown that cosmological parameters can be 
accurately measured by DECIGO and BBO with a precision of $\sim1\%$, 
assuming a large number of neutron star (NS) binaries, $\sim 10^6$.

In this paper, we show that the space-based GW detectors can also 
measure the Hubble parameter $H(z)$ from the GW standard sirens \footnote{Strictly speaking, this is true for flat geometry of the universe. For closed and open universe,   $H(z)$ is replaced with $H(z) \cos^{-1} \left[ \int_{0}^{z} \frac{dz^{'}}{H(z^{'})} \right]$, $H(z) \cosh^{-1} \left[ \int_{0}^{z} \frac{dz^{'}}{H(z^{'})} \right]$, respectively.}. In this method, the measured quantity is independent of that in the usual method of a standard siren, in which an observable is the luminosity distance as an integrated quantity of $H^{-1}(z)$. The method to measure the 
Hubble parameter at each redshift has been proposed by 
Bonvin {\it{et al.}} \cite{bib8,bib9}, who originally developed 
this idea in the observation of distant type Ia supernovae.  
In general, a large number of samples is necessary for the 
accurate measurement of $H(z)$, and they concluded that 
$10^5$ - $10^6$ supernovae are needed to achieve a few percent accuracy. 
In contrast to the supernovae observation which requires unrealistically 
large number of the samples and suffers from relatively large systematics, DECIGO and BBO are expected to detect a million of NS binaries with smaller systematic errors. Thus, the measurement of the Hubble parameter at high redshifts becomes even more feasible with standard sirens.  

The Hubble parameter $H(z)$ can be also measured by 
estimating the differential age of the oldest galaxies in 
each redshift bin \cite{bib24} and using the baryon acoustic oscillation (BAO) 
along the line-of-sight direction from the spectroscopic galaxy samples 
(e.g., \cite{bib32,bib40}). In this respect, 
the present method with GW observation is complementary, and 
useful for an independent cross check. 

This paper is organized as follows. In Sec.~\ref{sec2}, we briefly review 
the basic idea to measure the Hubble parameter via dipole of the luminosity 
distance $d_L$, originally proposed by Bonvin {\it{et al.}} \cite{bib8,bib9}, 
and derive the basic equations to estimate the accuracy of the 
Hubble parameter. In Sec.~\ref{sec3}, the method is applied to the GW 
observation. We briefly describe the GW standard siren and calculate the 
measurement accuracy of the Hubble parameter. 
In Sec.~\ref{sec:systematics}, systematic errrors are discussed and compared with the 
uncertainty coming from instrumental noise in GW observation. Finally, 
Sec.~\ref{sec:summary} gives a brief summary and discussions on the 
feasibility of our method. Throughout the paper, we adopt units 
$c=G=1$, and assume a flat universe.

%%%%%%%%%%%%%%%%%%%%%%%%%%%%%%%%%%%%%%%%%%%%%%%%%%%%%%%%%%%%%%%%%%%%
%%%%%%%%%%%%%%%%%%%%%%%%%%%%%%%%%%%%%%%%%%%%%%%%%%%%%%%%%%%%%%%%%%%%
\section{Hubble parameter from the dipole anisotropy of 
luminosity distance}
\label{sec2}
%%%%%%%%%%%%%%%%%%%%%%%%%%%%%%%%%%%%%%%%%%%%%%%%%%%%%%%%%%%%%%%%%%%%
%%%%%%%%%%%%%%%%%%%%%%%%%%%%%%%%%%%%%%%%%%%%%%%%%%%%%%%%%%%%%%%%%%%%

Consider the luminosity distance to some astronomical objects 
measured at redshift $z$ and angular position $\mathbf{n}$. 
In principle, the observations of many objects over the sky enables us 
to map out the angular distribution of luminosity distance, 
$d_L(z,\mathbf{n})$, and no directional dependence appears if the observer 
is at cosmological rest-frame (i.e., CMB rest-frame) in a homogeneous universe. 
However, there certainly exist tiny anisotropies in $d_L$ arising from 
the matter inhomogeneities of the large-scale structure and/or the local 
motion of the observer \cite{bib55}. As it has been shown in Ref.~\cite{bib9}, 
the dominant component of anisotropies is the dipole induced by the peculiar 
velocity of the observer, and the contribution to the higher multipoles 
coming from the weak gravitational lensing effect is basically small
\footnote{Nevertheless, the gravitational lensing magnification might be 
statistically measurable and has been recently discussed in 
Ref.~\cite{bib6}}. Then, we can expand the luminosity distance as 
%%%%%%%%%%%%%%%%%%%%%%%%%%%%%%%%%%%%%%%%%%%%%%%%%%%%%%%%%%%%%%%%%%%%
\begin{align}
d_L(z,\mathbf{n}) &= d_L^{(0)}(z) + d_L^{(1)}(z) \cos \theta\,\, ;
\label{eq:dL_expansion} \\
&d_L^{(0)}(z) \equiv  \frac{1}{4 \pi} \int d \mathbf{n} \, 
d_L(z,\mathbf{n}), \nonumber \\
&d_L^{(1)}(z) \equiv  \frac{3}{4 \pi} \int d \mathbf{n} \, 
(\mathbf{n} \cdot \mathbf{e}) d_L(z,\mathbf{n}), 
\label{eq:dL1}
\end{align} 
%%%%%%%%%%%%%%%%%%%%%%%%%%%%%%%%%%%%%%%%%%%%%%%%%%%%%%%%%%%%%%%%%%%%
where we define $\cos \theta=\mathbf{n}\cdot \mathbf{e}$, and 
the quantity $\mathbf{e}$ is the unit vector directed toward the dipole.

In the expression (\ref{eq:dL_expansion}), the first term in the right-hand 
side is the direction-averaged luminosity distance, which is identified 
with the one defined in the homogeneous and isotropic universe: 
%%%%%%%%%%%%%%%%%%%%%%%%%%%%%%%%%%%%%%%%%%%%%%%%%%%%%%%%%%%%%%%%%%%%
\begin{align}
d_L^{(0)}(z) &= (1+z) \int_0^{z} \frac{dz^{\prime}}{H(z^{\prime})}\,; 
\label{eq:dL0}
\end{align}
%%%%%%%%%%%%%%%%%%%%%%%%%%%%%%%%%%%%%%%%%%%%%%%%%%%%%%%%%%%%%%%%%%%%
with the Hubble parameter given by 
%%%%%%%%%%%%%%%%%%%%%%%%%%%%%%%%%%%%%%%%%%%%%%%%%%%%%%%%%%%%%%%%%%%%
\begin{align}
H(z) &= H_0 \left\{ \Omega_m (1+z)^3 + (1-\Omega_m ) (1+z)^{3(1+w_0+w_a)} 
\right. \nonumber \\
& \times \left. \exp \left[ -3 w_a \frac{z}{1+z} \right] \right\}^{1/2}.
\label{eq:Hubble} 
\end{align}
%%%%%%%%%%%%%%%%%%%%%%%%%%%%%%%%%%%%%%%%%%%%%%%%%%%%%%%%%%%%%%%%%%%%
Here we assumed the flat universe, and 
the dark energy equation-of-state parameter $w(a)=P/\rho$ 
parametrized by $w(a)=w_0+w_a (1-a)$. For later analysis of the error 
estimation, we adopt a fiducial set of cosmological parameters: 
$\Omega_m=0.3$, $w_0=-1$, $w_a=0$, and $H_0=72$\,km s$^{-1}$\,Mpc$^{-1}$.

On the other hand, the second term on the right-hand-side of 
Eq.~(\ref{eq:dL_expansion}) arises from the Doppler effect due to the motion of the observer. To derive the expression for $d_L^{(1)}$, 
we approximate the propagation of GW or light follows the 
trajectory of a null geodesic. Writing  
the luminosity distance to an astronomical object as a function of 
the conformal time $\eta$ at which the source emits GW or light, 
the Doppler effect leads to \cite{bib8,bib9}
%%%%%%%%%%%%%%%%%%%%%%%%%%%%%%%%%%%%%%%%%%%%%%%%%%%%%%%%%%%%%%%%%%%%
\begin{align}
d_L(\eta,\mathbf{n}) = d_L^{(0)}(\eta) 
[ 1-\mathbf{n} \cdot \mathbf{v}_0 ]\;,
\label{eq:dL_Doppler}
\end{align}
%%%%%%%%%%%%%%%%%%%%%%%%%%%%%%%%%%%%%%%%%%%%%%%%%%%%%%%%%%%%%%%%%%%%
where the vector $\mathbf{v}_0$ indicates the peculiar velocity of a
local observer, which is small enough relative to the light velocity. 
Note that the motion of the local observer also induces the Doppler effect 
in the redshift, $z=\overline{z}+\delta z$, where the unperturbed redshift $\bar{z}$ is defined as $\bar{z}=a^{-1}(\eta)-1$.
To first order in $\mathbf{v}_0$, we have
%%%%%%%%%%%%%%%%%%%%%%%%%%%%%%%%%%%%%%%%%%%%%%%%%%%%%%%%%%%%%%%%%%%%
\begin{align}
\delta z = - (\mathbf{n} \cdot \mathbf{v}_0) (1+\overline{z}).  
\label{eq:delta_z}
\end{align}
%%%%%%%%%%%%%%%%%%%%%%%%%%%%%%%%%%%%%%%%%%%%%%%%%%%%%%%%%%%%%%%%%%%%
Rewriting Eq. (\ref{eq:dL_Doppler}) in 
terms of the redshift $z$, then expanding $d_L^{(0)}(z-\delta z)$ up to the first order, and using Eqs.\,(\ref{eq:dL0}) and (\ref{eq:delta_z}) give
\begin{eqnarray}
d_L(z,\mathbf{n}) &=& \left[ d_L^{(0)}(z)- \frac{\partial d_L^{(0)}(z)}{\partial z} \delta z \right] [ 1-\mathbf{n} \cdot \mathbf{v}_0 ] \nonumber \\
&=& d_L^{(0)}(z)+\frac{(1+z)^2}{H(z)}\,
(\mathbf{n}\cdot\mathbf{v}_0) \;.
\end{eqnarray}
Comparing the above expression with Eqs.(\ref{eq:dL_expansion}) and 
(\ref{eq:dL1}), we arrive at 
%%%%%%%%%%%%%%%%%%%%%%%%%%%%%%%%%%%%%%%%%%%%%%%%%%%%%%%%%%%%%%%%%%%%
\begin{align}
d_L^{(1)}(z)= \frac{|\mathbf{v}_0|(1+z)^2}{H(z)} 
\label{eq:dL1_Doppler}
\end{align}
%%%%%%%%%%%%%%%%%%%%%%%%%%%%%%%%%%%%%%%%%%%%%%%%%%%%%%%%%%%%%%%%%%%%
with the direction of the dipole specifically chosen as 
$\mathbf{e}=\mathbf{v}_0/|\mathbf{v}_0|$. In the expression 
(\ref{eq:dL1_Doppler}), the magnitude of dipole anisotropy is 
inversely proportional to the 
Hubble parameter at the source redshift, because the perturbed luminosity distance corresponds to the derivative of Eq.\,(\ref{eq:dL0}). Recalling the fact that  
the motion of the local observer also induces the same size of 
dipole anisotropy in the CMB and its amplitude is estimated as 
$|\mathbf{v}_0|=369.1\pm0.9$km s$^{-1}$ \cite{bib13}. Then the dipole anisotropy in the luminosity distance to 
high-$z$ objects gives a direct measure of $H(z)$. 

\begin{figure}[t]
\begin{center}
\includegraphics[width=8cm]{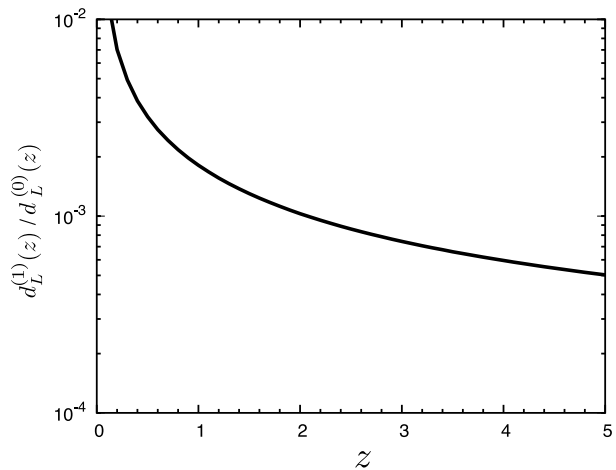}
\caption{Ratio of $d_L^{(1)} (z)$ to $d_L^{(0)} (z)$, in which the CMB 
dipole of $|v_0| = 369.1 \pm 0.9 \,{\rm{km}}/ {\rm{sec}}$ measured by 
Wilkinson Microwave Anisotropy Probe (WMAP) \cite{bib13} is used.}
\label{fig:ratio_dL1_dL0}
\end{center}
\end{figure}

Now, let us discuss the statistical error in the measurement of the
Hubble parameter. We add the measurement error of the 
luminosity distance $\delta d_L(z,\mathbf{n})$ into 
Eq.~(\ref{eq:dL_expansion}):  
\begin{align}
d_L(z,\mathbf{n}) = d_L^{(0)} (z)+ d_L^{(1)}(z) (\mathbf{n} \cdot \mathbf{e}) 
+\delta d_L(z,\mathbf{n}) \;. \nonumber
\end{align} 
Then the measurement error of $d_L^{(1)}$ is estimated from 
the definition of $d_L^{(1)}$, and is expressed as
\begin{align}
\delta d_L^{(1)}(z) = \frac{3}{4 \pi} \int d \mathbf{n} \, (\mathbf{e} 
\cdot \mathbf{n}) \delta d_L(z,\mathbf{n}) \;. \nonumber
\end{align}
The variance of this becomes
\begin{align}
\left[ \Delta d_L^{(1)} (z) \right]^2 &\equiv \left\langle 
\left[ \delta d_L^{(1)} (z) \right]^2 \right\rangle 
\nonumber \\
&= \left( \frac{3}{4\pi} \right)^2 \int d\mathbf{n} \int 
d\mathbf{n}^{\prime}\, 
\nonumber \\
& \times (\mathbf{e} \cdot \mathbf{n}) (\mathbf{e} \cdot 
\mathbf{n}^{\prime}) \langle \delta d_L(z,\mathbf{n}) \delta 
d_L(z,\mathbf{n}^{\prime}) \rangle \;, 
\nonumber \\
& \label{eq:var_dL1} 
\end{align}
where the bracket represents the ensemble average over the sources 
given at $z$. Assuming that the distance errors measured from each GW 
source are statistically independent and isotropic, we have
\begin{align}
\langle \delta d_L(z,\mathbf{n}) \delta d_L(z,\mathbf{n}^{\prime}) 
\rangle = 4\pi \left[ \Delta d_L^{(0)}(z) \right]^2 \delta^2 
(\mathbf{n}-\mathbf{n}^{\prime}) 
\nonumber
\end{align} 
where the quantity $\left[ \Delta d_L^{(0)}(z) \right]^2$ is the variance 
of $d_L^{(0)}$. Substituting this into Eq.~(\ref{eq:var_dL1}), we obtain
\begin{align}
\left[ \Delta d_L^{(1)} (z) \right]^2 &= \frac{9}{4\pi} 
\left[ \Delta d_L^{(0)}(z) \right]^2 \int d\mathbf{n}\, 
(\mathbf{e} \cdot \mathbf{n})^2 \;,  
\nonumber \\
&= 3 \left[ \Delta d_L^{(0)}(z) \right]^2 \;. 
\label{eq:var_dL1_rev}  
\end{align}
From this and the expression (\ref{eq:dL1_Doppler}), the measurement 
error of the Hubble parameter at a given redshift $z$ is related to the 
error of the direction-averaged luminosity distance, $\Delta d_L^{(0)}(z)$, 
and is given by 
\begin{align}
\frac{\Delta H(z)}{H(z)} &= \frac{\Delta d_L^{(1)} (z)}{d_L^{(1)} (z)} 
\nonumber \\
&= \sqrt{3} \left[ \frac{d_L^{(1)} (z)}{d_L^{(0)} (z)} \right]^{-1} 
\left[ \frac{\Delta d_L^{(0)} (z)}{d_L^{(0)} (z)} \right]  \;. 
\label{eq:error_H}
\end{align}
Here we ignored the velocity error in CMB observation. In fact, the contribution is negligible as we will show in Sec.\,\ref{sec:systematics}. The factor $[d_L^{(1)}/d_L^{(0)}]^{-1}$ 
is typically very large in the present case with $|\mathbf{v}_0|\ll1$. 
In Fig.~\ref{fig:ratio_dL1_dL0}, we plot the 
ratio of the dipole to the monopole in the luminosity distance. Figure
implies that even a negligibly small error in the averaged distance 
$d_L^{(0)}$ can produce a large scatter in $H(z)$.

Note that the mean error on the Hubble parameter is reduced to 
$\Delta H/\sqrt{N}$ if we observe $N$ independent sources at a given 
redshift bin. Thus, for a precision measurement of Hubble parameter, 
we need not only an accurate determination of averaged distance to each source, 
$\Delta d_L^{(0)}/d_L^{(0)}\ll1$ but also a large number of sources. In the case of the type Ia supernovae as standard candles, the averaged 
distance error is related to the intrinsic magnitude error $\Delta m$ as 
$\Delta d_L^{(0)}/d_L^{(0)}=(\ln10\,/5)\Delta m$. Adopting 
an optimistic value $\Delta m=0.1$, it yields the error 
$\Delta d_L^{(0)}/d_L^{(0)}\approx0.05$, but 
to achieve a few percent accuracy in the Hubble parameter,  
we need unrealistically large number of samples of the order 
$10^6$. In other words, for a reasonable number of $10^4$ samples (see e.g. \cite{bib53,bib54}),  
the systematics in the averaged distance should be reduced to 
$\Delta m<0.01$, which seems very difficult from the empirical 
calibration. 

%%%%%%%%%%%%%%%%%%%%%%%%%%%%%%%%%%%%%%%%%%%%%%%

\section{GW standard sirens}
\label{sec3}

In this section, we consider the GW standard sirens as an alternative 
probe to measure the Hubble parameter from the dipole anisotropy of 
luminosity distance.  The advantage to use the standard sirens observed 
by space-based GW detectors is that the expected number of sources 
(NS binaries) is much larger than that of the type Ia supernovae, 
and the sources are distributed deeply enough at higher redshifts. 
Moreover, the NS binary would be a clean GW source, and with 
DECIGO or BBO, the luminosity distance $d_L(z,\mathbf{n})$ can be 
accurately measured with less systematics.

%%%---%%---%%---%%---%%---%%---%%---%%---%%---%%---%%---%%---%%---%%
\subsection{Luminosity distance error}
\label{subsec:luminosity_distance}

Let us first estimate the distance error 
of the standard siren, taking account of the instrumental 
noise of the GW detector. Possible systematic errors will be discussed 
later. In GW experiments, a direct observable is the 
waveform of the GW signal, and comparing it with a theoretical template, we not only determine the system parameters of GW source but also 
extract the cosmological information. 

For a single binary system, the Fourier transform of the GW waveform is 
expressed as a function of frequency $f$ \cite{bib46,bib47},
\begin{equation}
\tilde{h} (f) = \frac{A}{d_L(z)} M_z^{5/6} f^{-7/6} e^{i \Psi(f)} \;, 
\label{eq:waveform}
\end{equation}
where $d_L$ is the luminosity distance, and  
the quantity $M_z=(1+z) \eta^{3/5} M_t$ is the 
redshifted chirp mass with the total mass $M_t=m_1+m_2$ and the symmetric 
mass ratio $\eta=m_1 m_2/M_t^2$. Here, 
the constant $A$ is given by $A= (\sqrt{6}\, \pi^{2/3})^{-1}$, 
which is multiplied by the factor $\sqrt{4/5}$ for a geometrical 
average over the inclination angle of a binary  
\footnote{For the geometric factor arising from the 
non-orthogonal detector arms ($60^\circ$), we incorporate this effect 
into the detector noise curve $P(f)$. The factor arising from 
detector's angular response is also taken into account in evaluating 
the noise curve by averaging over the sky.}.
The function $\Psi(f)$ represents the frequency-dependent phase arising from 
the orbital evolution, and at the order of the {\it{restricted}} $1.5$ 
post-Newtonian (PN) approximation, it is given by \cite{bib46,bib47} 
\begin{eqnarray}
\Psi (f) &=& 2\pi f \,t_c -\phi_c -\frac{\pi}{4} +\frac{3}{128} 
(\pi M_z f)^{-5/3} \nonumber \\
&& \times \left[ 1+ \frac{20}{9} \left(\frac{743}{336} + 
\frac{11}{4} \eta \right)
\eta^{-2/5} (\pi M_z f)^{2/3} \right. \nonumber \\
&& \left. - 16 \pi \eta^{-3/5} (\pi M_z f) \right] \;, 
\label{eq:phase_1.5PN}
\end{eqnarray}
where $t_c$ and $\phi_c$ are the time and phase at coalescence, 
respectively. The first term in the bracket in Eq.~(\ref{eq:phase_1.5PN}) 
corresponds to Newtonian-order dynamics 
and the other remaining terms represent the Post-Newtonian 
order corrections in powers of 
$v \sim (\pi M_z f)^{1/3}$. In principle, there additionally appears a phase correction due to cosmic expansion, and  
the Hubble parameter $H(z)$ can be also measured from this term 
\cite{bib18,bib5}. Although the inclusion of the phase correction slightly changes the size of the errors in binary parameters, it does not seriously affect the estimation of the luminosity distance $d_L$. In addition, the sensitivity of the phase correction to the Hubble parameter is rather small. Thus, we may safely ignore the phase correction due to cosmic expantion in the subsequent analysis. 

In Eqs.~(\ref{eq:waveform}) and (\ref{eq:phase_1.5PN}) , 
there are five unknown parameters to be determined observationally, i.e.,  
$M_z$, $\eta$, $t_c$, $\phi_c$, and $d_L$. Except for the luminosity distance, 
the four parameters merely carry the information on the individual 
property of the binary system. For simplicity, we consider 
the equal-mass NS binaries with $1.4 M_{\odot}$, which lead to 
$M_z=1.22(1+z) M_{\odot}$ and $\eta=1/4$, and set the other parameters to 
$t_c=0$ and $\phi_c=0$. 

Since the GW observation can only determine 
the redshifted chirp mass $M_z$, the redshift of each binary has to be 
measured from an electromagnetic counterpart. 
According to Cutler and Holz \cite{bib6}, the angular resolution of BBO is 
$\sim 1-100\,{\rm{arcsec}}^2$, with which we can identify the host 
galaxy of the binary. We thus suppose that the redshift of any binary system 
is obtained from the electromagnetic observations. Note that the Doppler 
effect by the local motion also affects the redshifted chirp 
mass, and the dipole anisotropy might be measured through the spatial 
distribution of the observed chirp mass if the intrinsic scatter in the 
mass distribution of NS binaries is very small. The feasibility to 
measure the dipole anisotropy from the chirp mass might be interesting, 
but we need a more detailed study on the formation history of NS binaries, 
and we here simply ignore this effect in the parameter estimation. 

\begin{figure}[t]
\begin{center}
\includegraphics[width=8.5cm]{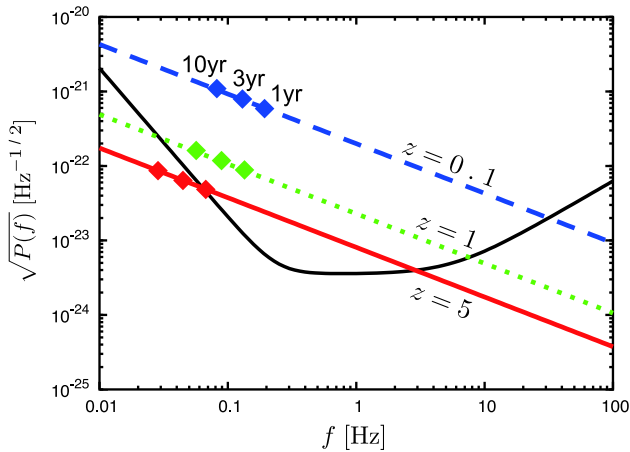}
\caption{Sky-averaged DECIGO noise curve. (Arm angle $60^{\circ}$ is taken 
into account.) Diagonal lines represent frequency evolutions of a NS-NS 
binary at $z=5$ (solid, red), $z=1$ (dotted, green), and $z=0.1$ (dashed, 
blue). Diamonds on the lines from the right to the left denote the frequency 
of the binary $1\,{\rm{yr}}$, $3\,{\rm{yr}}$, and $10\,{\rm{yr}}$ before the 
merger.}
\label{fig:noise_curve}
\end{center}
\end{figure}

The fundamental basis to estimate the distance error for a single binary 
is the Fisher matrix formalism. The Fisher matrix for a single binary 
is given by \cite{bib46,bib48}
\begin{eqnarray}
\Gamma_{ab}&=& 4 \sum_{i=1}^{8}\, {\rm{Re}} \int_{f_{\rm{min}}}^{f_{\rm{max}}}
 \frac{\partial_{a} \tilde{h}_{(i)}^{\ast}(f)\, \partial_{b}
 \tilde{h}_{(i)}(f)}{P(f)} df \;, 
\label{eq3}
\end{eqnarray}
where $\partial_a$ denotes a derivative with respect to a parameter $\theta_a$; 
$M_z$, $\eta$, $t_c$, $\phi_c$, and $d_L$. The quantity $\tilde{h}_{(i)}$ 
represents the GW signal obtained from the $i$-th interferometer. 
Since two independent signals are obtained for each cluster 
\cite{bib21}, DECIGO has the eight interferometric signals 
in total, each of which is supposed to have an identical 
detector response and noise power spectrum $P(f)$. 
The analytical fit of noise spectrum \footnote{The 
noise power spectrum presented here is slightly different from the  
one given by Ref.~\cite{bib7}, who considered orthogonal 
detector arms and optimal incidence of a GW. } is given by
\begin{eqnarray}
P (f)&=& 4.21\times 10^{-50} \biggl(\frac{f}{1\rm{Hz}} \biggr)^{-4} 
+1.25\times 10^{-47} \; \nonumber \\
&& +3.92\times 10^{-49} \biggl(\frac{f}{1\rm{Hz}} \biggr)^{2} \; 
\rm{Hz}^{-1} \;. \nonumber
\end{eqnarray}
In Fig.~\ref{fig:noise_curve}, the noise spectrum of DECIGO is shown,  
together with the evolutionary tracks of the NS binary located at 
three different redshifts, $z=0.1$, $1$, and $5$. In each track, 
the symbols indicate the frequency at the $10$, $3$ and $1$ years 
before the time of binary coalescence (from left to right). In this 
respect, the lower cutoff of the frequency $f_{\rm min}$ 
should be incorporated into the integration in 
Eq.~(\ref{eq3}),  and is given by the function of observation 
time $T_{\rm obs}$ as well as the redshift and mass: 
\begin{equation}
f_{\rm{min}} = 0.233 \left( \frac{1M_{\odot}}{M_z} \right)^{5/8} 
\left( \frac{1\,{\rm{yr}}}{T_{\rm{obs}}} \right)^{3/8} \; {\rm{Hz}}\;. 
\label{eq:f_min}
\end{equation}
Note that the coalescence frequency of the NS binary is 
typically $\sim {\rm{kHz}}$, and thus the upper cutoff of the 
frequency naturally arises from the noise curve. For the computational 
purpose, we set $f_{\rm{max}}=100$\,Hz. 

\begin{figure}[t]
\begin{center}
\includegraphics[width=8cm]{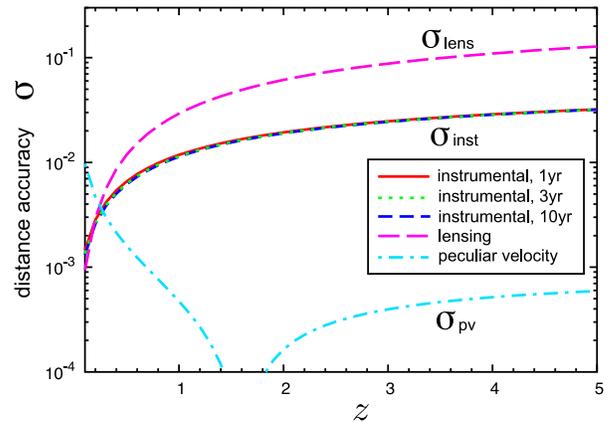}
\caption{Measurement accuracy of the luminosity distance with a single binary as a function of redshifts. The curves tagged $\sigma_{\rm{inst}}$ are those determined only by instrumental noise and with the observation time $1\,{\rm{yr}}$ (red, solid curve), $3\,{\rm{yr}}$ (green, dotted curve), and $10\,{\rm{yr}}$ (blue, short-dashed curve), respectively. The lensing error and the peculiar velocity error are represented by magenta (long-dashed) and light blue (dot-dashed) curves.}
\label{fig:distance_accuracy}
\end{center}
\end{figure}

Given the numerically evaluated Fisher matrix, 
the marginalized 1-sigma error of a parameter, $\Delta\theta_a$ 
is estimated from the inverse Fisher matrix 
\begin{align}
\Delta \theta_a = \sqrt{\{\mathbf{\Gamma}^{-1}\}_{aa}}. 
\end{align}
In Fig.~\ref{fig:distance_accuracy}, 
the resultant error of the luminosity distance for a single binary, 
$\sigma_{\rm{inst}}$, is plotted against a source redshift, assuming the observation time $1\,{\rm{yr}}$ (solid curve), $3\,{\rm{yr}}$ (dotted curve), and $10\,{\rm{yr}}$ (dashed curve). The overlap of these three curves indicates that $\sigma_{\rm{inst}}$ hardly depends on the observation time, because the observation time appears only through the cutoff frequency $f_{\rm{min}}$ with the fractional power of $3/8$ and the improvement of the precision is generally slow. Even for a single binary system, the precision of a few percent level is easily achievable for the distance measurement in the absence of systematic errors, and this is also true for a rather high-$z$ binary.  

%%%---%%---%%---%%---%%---%%---%%---%%---%%---%%---%%---%%---%%---%%
\subsection{Accuracy of Hubble parameter}
\label{subsec:Hubble}

Given the uncertainty of the averaged luminosity distance 
for each binary, the accuracy of the Hubble parameter 
is estimated from Eq.~(\ref{eq:error_H}), and with the ensemble 
over the $\Delta N(z)$ independent binary systems 
in the vicinity of the redshift $z$, we can get an improved constraint 
on the Hubble parameter at each redshift bin.

Here, to derive the measurement error of the Hubble parameter, 
we adopt the following fitting form of the NS-NS merger rate 
given by Ref.~\cite{bib11}:  
%%%%%%%%%%%%%%%%%%%%%%%%%%%%%%%%%%%%%%%%%%%%%%%%%%%%%%%%%%%%%%%%%%%%
\begin{align}
\dot{n}(z) &= \dot{n}_0\, r(z)\,\, ;\quad
r(z) = \left\{        
\begin{array}{ll}
1+2 z  & (z \leq 1) \\
\frac{3}{4} (5-z) &(1 < z \leq 5) \\
 0 & (5 < z)
\end{array}
\right. \;, 
\label{eq:merger_rate}
\end{align}
%%%%%%%%%%%%%%%%%%%%%%%%%%%%%%%%%%%%%%%%%%%%%%%%%%%%%%%%%%%%%%%%%%%%
where the function $r(z)$ is estimated based on the star formation history 
inferred from the UV luminosity \cite{bib10}. 
The quantity $\dot{n}_0$ represents the merger rate at present. 
Though it is still uncertain, we adopt the most recent estimate, 
$\dot{n}_0=10^{-6}\,{\rm{Mpc}}^{-3}\, {\rm{yr}}^{-1}$, as 
a reliable and confident estimate based on extrapolations from the 
observed binary pulsars in our Galaxy \cite{bib12}. Then, the number of 
NS binaries in the redshift interval $[z-\Delta z/2,z+\Delta z/2]$ observed 
during $T_{\rm obs}$, $\Delta N(z)$, is given by \cite{bib11}
%%%%%%%%%%%%%%%%%%%%%%%%%%%%%%%%%%%%%%%%%%%%%%%%%%%%%%%%%%%%%%%%%%%%
\begin{align}
\Delta N(z)=T_{\rm obs}\,\int_{z-\Delta z/2}^{z+\Delta z/2}
dV_c(z') \,\,\frac{\dot{n}(z')}{1+z'} \;,
\label{eq:ns-merger-rate}
\end{align}
%%%%%%%%%%%%%%%%%%%%%%%%%%%%%%%%%%%%%%%%%%%%%%%%%%%%%%%%%%%%%%%%%%%%
where $dV_c$ means the comoving volume element defined as 
$dV_c(z)=4\pi r^2(z) dz/H(z)$ with the comoving radial distance 
$r(z)=d_L(z)/(1+z)$.

\begin{figure}[t]
\begin{center}
\includegraphics[width=8cm]{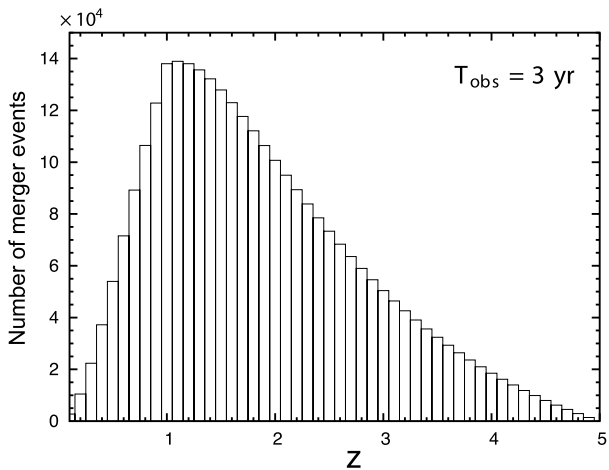}
\caption{Number of NS-NS binaries (in the unit of $10^4$) that would be observed by DECIGO in each redshift bin of $\Delta z =0.1$ at a redshift $z$ during $3\,{\rm{yr}}$ observation. 
As is manifest from Eq.\,(\ref{eq:ns-merger-rate}), the number of the binaries scales linearly with $T_{\rm{obs}}$.} 
\label{fig:NS_binary_dist}
\end{center}
\end{figure}

\begin{figure}[t]
\begin{center}
\includegraphics[width=8cm]{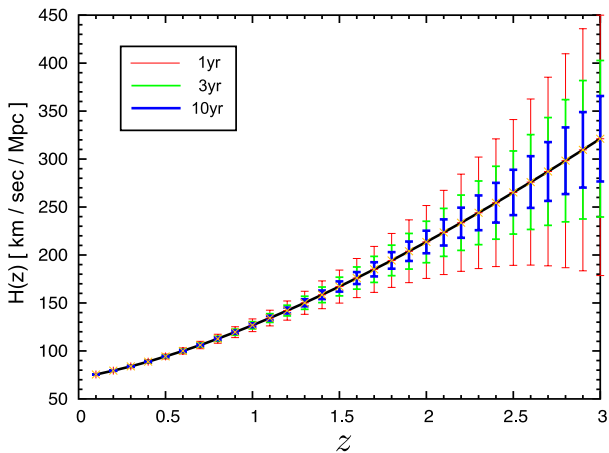}
\caption{The Hubble parameter calculated with our fiducial cosmological 
parameters (solid curve) and $1\sigma$-error bars estimated in the cases 
that we use all binaries observed by DECIGO during the observation time, 
$1\,{\rm{yr}}$ (red), $3\,{\rm{yr}}$ (green), and $10\,{\rm{yr}}$ (blue) (long observation time corresponds to the smaller error bar).}
\label{fig:Hubble_diagram}
\end{center}
\end{figure}

In Fig.~\ref{fig:NS_binary_dist},  
observed redshift distribution of NS binaries $\Delta N(z)$ is plotted, 
assuming the $3$ year observation and the redshift width $\Delta z=0.1$. 
The total number of NS binaries is $\sim10^6$, which is much larger 
than the expected number of type Ia supernovae. Note that 
the number of merger events increases with $T_{\rm obs}$, and thus the 
accuracy of the Hubble parameter is improved by a factor 
$T_{\rm obs}^{1/2}$. Combining this and the distance error in previous 
subsection, Fig.~\ref{fig:Hubble_diagram} shows the expected errors 
for the Hubble parameter measured from the dipole anisotropy. 
The three different error bars in each redshift bin represent the results 
from the $1$-, $3$-, and $10$-year observations (from large to small 
sizes). The Figure implies that up to the redshift $z=1$, the Hubble parameter 
can be accurately measured with a precision of $2-5\%$, $1-3\%$, and 
$0.7-1.5\%$ for the observation time of $1$, $3$, and $10$ years, respectively. Even at $z=2$, the Hubble parameter 
can be measured with a precision of $18\%$, $10\%$, and 
$6\%$ for the observation time of $1$, $3$, and $10$ years, respectively. 
This is quite impressive in the sense that a GW standard siren has a nearly equal sensitivity to the Hubble parameter with other complementary methods 
such as BAO. Another noticeable point using the standard sirens is 
that we can trace the redshift evolution of Hubble parameter even at 
higher redshift $z\gtrsim1$. Although the number of high-$z$ NS binaries 
is highly uncertain, the standard sirens would be potentially powerful to 
probe the early-time cosmic expansion, and should deserve further 
investigation.  

%%%%%%%%%%%%%%%%%%%%%%%%%%%%%%%%%%%%%%%%%%%%%%%
\section{Systematic errors}
\label{sec:systematics}

So far, we have discussed the accuracy of Hubble parameter 
taking only account of the distance error associated with the instrumental noise. However, there are several effects which may systematically affect 
the measurement of dipole anisotropies in the luminosity distance, 
leading to increasing the error in the Hubble parameter. 
Among them, a dominant contribution 
may come from the gravitational lensing magnification induced by the matter 
inhomogeneities of large-scale structure along the line of sight 
(e.g., \cite{bib51,bib14,bib50,bib49}), which systematically changes the luminosity distance to each binary system. 
Another important effect would be the peculiar velocity of the binary along the line of sight, which randomly contributes to measurement error via Doppler effect. 
These systematic errors to the averaged luminosity distance are summarized as 
\begin{equation}
\left[ \frac{\Delta d_L^{(0)}(z)}{d_L^{(0)}(z)} \right]^2 = 
\sigma_{\rm{inst}}^2(z) + \sigma_{\rm{lens}}^2(z) + \sigma_{\rm{pv}}^2(z) \;, 
\label{eq:systematics_in_dL}
\end{equation}
where $\sigma_{\rm inst}$ is the error associated with the GW experiment 
in Sec.~\ref{subsec:luminosity_distance}, $\sigma_{\rm lens}$ is the lensing error, and $\sigma_{\rm pv}$ is the peculiar-velocity error.

There are several studies on the effect of lensing magnification, 
particularly focusing on 
the distance measurement from the type Ia supernovae.  
Holz and Linder \cite{bib14} estimated the lensing error on the 
distance measurement by using Monte Carlo simulation, and 
assuming the Gaussian form of lensing magnification probability,  
they derived a fitting formula for the systematic error. 
Later, the significance of non-Gaussian tail has been recognized 
\cite{bib15,bib16}, and it turned out that 
this effect reduces the lensing error by a factor of 
$1.5$ - $2$, compared to the Gaussian distribution. More recently, 
Hirata, Holz, and Cutler \cite{bib15} adopted a log-normal distribution 
for the magnification probability and 
obtained the fitting formula for the (averaged) distance error:  
\begin{equation}
\sigma_{\rm{lens}}(z)=0.066 \left[ \frac{1-(1+z)^{-0.25}}{0.25} \right]^{1.8} 
\;. \label{eq:error_lensing}
\end{equation}
In what follows, we adopt the lensing error in Eq.~(\ref{eq:error_lensing}).

As for the peculiar velocity error, the clustering of galaxies induced by the gravity leads to a coherent and/or virialized random motion, which gives rise to the 
Doppler effect and affects the determination of cosmological redshift 
via the spectroscopic measurement. In addition, binary barycentric motion itself in the host galaxy also leads to the Doppler effect, 
which causes random fluctuations in the luminosity distance. These two systematic effects can be of the same order and can be translated into the distance error as \cite{bib17}
\begin{align}
\sigma_{\rm{pv}}(z) = \left| 1-\frac{(1+z)^2}{H(z)d_L(z)} \right| 
\sigma_{\rm{v,gal}} \;. \nonumber
\end{align}
Here, $\sigma_{\rm{v,gal}}$ is the one-dimensional velocity dispersion of the galaxy. Taking account of the non-linear effect of gravity, it often set to 
$\sigma_{\rm{v,gal}}=300$km\,s$^{-1}$, mostly independent of the redshifts 
\cite{bib52}. 

In Fig.~\ref{fig:distance_accuracy}, the distance errors for a 
single binary system from the lensing magnification, $\sigma_{\rm{lens}}$, and peculiar 
velocity, $\sigma_{\rm{pv}}$, are overlaid, together with the uncertainty from the GW 
observations. It turns out that the lensing magnification could 
dominate the distance error, and exceeds the error from the GW 
experiments at $z\gtrsim0.2$. Thus, the lensing magnification 
could be potentially a main source of the distance error, and 
the accuracy of the Hubble parameter might be somewhat degraded.  

%%%%%%%%%%%%%%%%%%%%%%%%%%%%%%%%%%%%%%%%%%%%%%%%%%%%%%%%%%%%%%%%%%%%
\begin{figure}[t]
\begin{center}
\includegraphics[width=8cm]{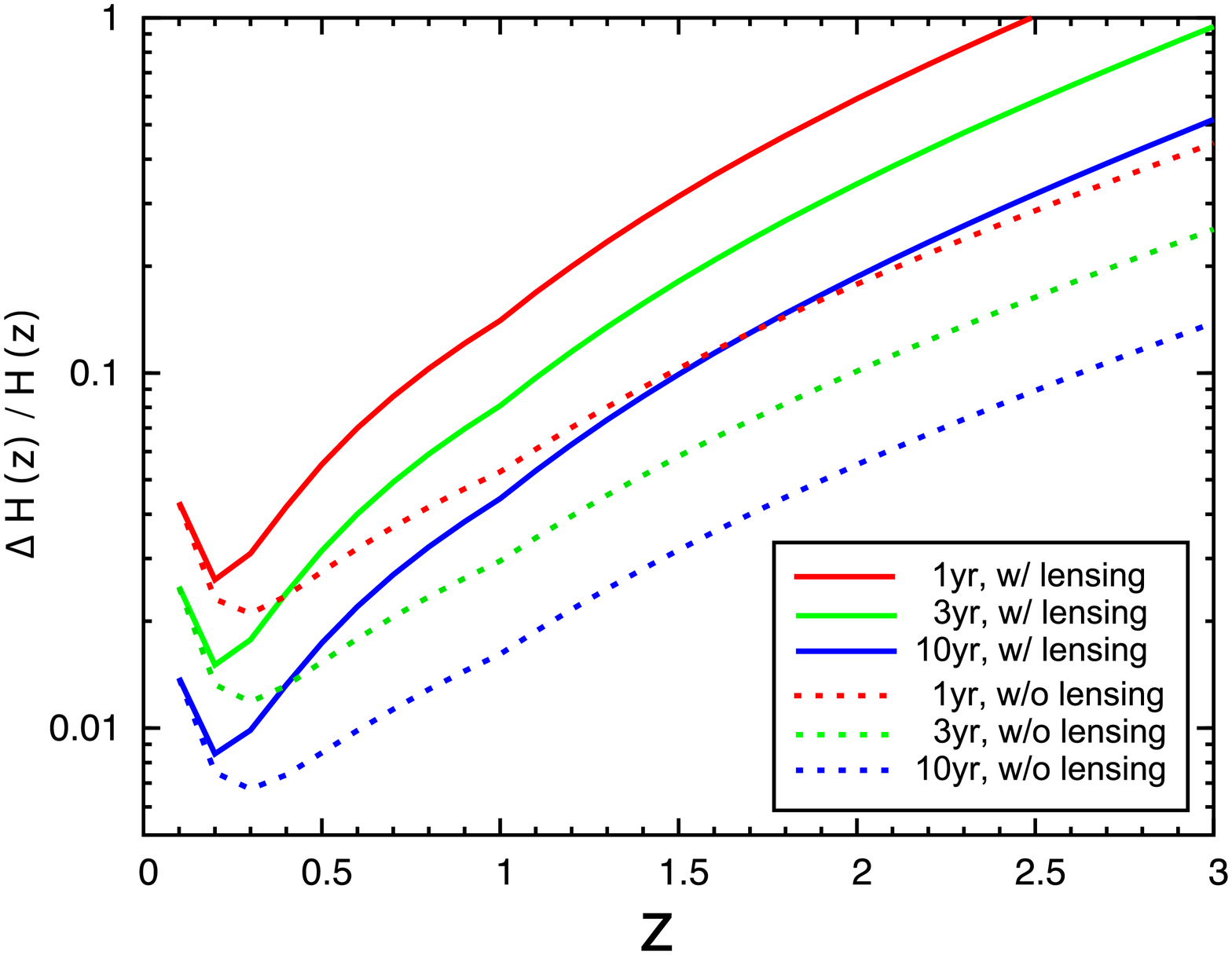}
\caption{Measurement accuracy of the Hubble parameter with all observed 
binaries. We plot the measurement accuracies, including the lensing error, 
with the observation time $T_{\rm{obs}}=1\,{\rm{yr}}$, $3\,{\rm{yr}}$, and 
$10\,{\rm{yr}}$ from the top to the bottom, respectively (solid curves), 
and those without the lensing error (dotted curves).}
\label{fig:Hubble_error}
\end{center}
\end{figure}   
%%%%%%%%%%%%%%%%%%%%%%%%%%%%%%%%%%%%%%%%%%%%%%%%%%%%%%%%%%%%%%%%%%%%

In Fig.~\ref{fig:Hubble_error}, 
the size of the measurement error for Hubble parameter, 
$\Delta H(z)/H(z)$, is plotted as function of redshift,  
taking account of all the systematics and the binary distribution. 
Comparing with the case without lensing error (dotted curves), 
the resultant accuracy of Hubble parameter are degraded. Nevertheless, even including the lensing systematics, the Hubble parameter up to $z\lesssim1$ can be 
accurately measured with the precision $3-14\,\%$, $1.5-8\,\%$, and 
$0.8-4\,\%$, for $T_{\rm{obs}}=1\,{\rm{yr}}$, $3\,{\rm{yr}}$, 
and $10\,{\rm{yr}}$, respectively. Although the lensing effect is 
potentially crucial for the cosmological application of standard sirens, 
the technique to reduce the lensing effect has been recently exploited \cite{bib15,bib16,bib43,bib44}, 
and the feasibility of the method has been discussed. With an improved 
technique developed near future, the lensing systematics would be 
removed, and one could approach the limit determined by the instrumental noise. 

We briefly comment on the uncertainty in the amplitude 
of dipole $\mathbf{v}_0$. The current constraint on the motion of local
observer comes from the CMB observation, and the estimated error of 
$\mathbf{v}_0$ directly affects the accuracy of Hubble parameter in Eq.\,(\ref{eq:error_H}). However, the current observation 
produces 
\begin{align}
\frac{\Delta H(z)}{H(z)}=
\frac{\Delta |\mathbf{v}_0|}{|\mathbf{v}_0|} \approx 2.44 \times 10^{-3}.   
\nonumber
\end{align}
Therefore, the systematics in the dipole from the CMB observation 
gives a tiny contribution, and can be ignored. 

Finally, note that the results obtained here rely on the fact that we can successfully
identify the redshifts of all host galaxies. In practice, the identification would be difficult as increasing the redshift, because galaxies at high-z are fainter and the time required for a spectroscopic measurement is much longer than that for low-z galaxies. The resultant accuracy of the Hubble parameter measurement would be degraded, being proportional to $1/\sqrt{N(z)}$, where $N(z)$ is the number of galaxies in a bin at the redshift z. Thus,  the measurement accuracy depends on the capability of galaxy redshift survey available in the
future.

Another important issue is detector calibration, which would potentially affect the measurement accuracy of the luminosity distance. It is rather crucial not only for the Hubble parameter measurment, but also for the subtraction of the neutron star binary foreground in order to achieve the detection of an inflationary gravitational-wave background as a primary science goal of DECIGO/BBO. Although this is beyond the scope of this paper, the issue should be considered seriously and addressed in the future.

%%%%%%%%%%%%%%%%%%%%%%%%%%%%%%%%%%%%%%%%%%%%%%%
\section{Summary}
\label{sec:summary}

In this paper, we have shown that the redshift evolution of 
Hubble parameter can be measured by utilizing a large number of NS binaries 
observed by space-based GW detectors such as DECIGO and BBO. Although this 
method requires a large number of samples $10^5$-$10^6$ to 
measure $H(z)$ with the accuracy of a few $\%$, DECIGO and BBO will detect enough number of samples up to high redshifts. 
Including the lensing magnification,  the Hubble parameter as a 
function of redshift up to $z=1$ can be determined with 
the accuracy of $3-14\,\%$, $1.5-8\,\%$, and $0.8-4\,\%$, 
for the observation time 
$T_{\rm{obs}}=1$, $3$, and $10\,{\rm{yr}}$, respectively. With 
a technique to remove the lensing magnification, we can further improve 
the accuracy by a factor of $\sim 2.7$ at $z=1$. Although the feasibility of the method, particularly on the redshift indentification of the host galaxies and the detector calibration, still need to be investigated in more detail, the present method puts forward an interesting possibility to enlarge the science with standard sirens as a by-product, and it is complementary to other methodologies to measure Hubble parameter. 

\begin{acknowledgments}
We acknowledge N. Kanda, T. Tanaka, K. Yagi, J. Yokoyama, and C. Yoo for 
helpful discussions and valuable comments. We also thank T.~Smith for his effort to improve our manuscript. A. N. and S. S. are supported 
by a Grant-in-Aid for JSPS Fellows. S. S. is also supported by JSPS through 
Excellent Young Researchers Overseas Visit Program. A. T. and S. S. is 
supported in part by a Grants-in-Aid for Scientific Research from 
the JSPS (No. 21740168 for A. T. and No. 21-00784 for S. S.).  
\end{acknowledgments}
 
%%%%%%%%%%%%%%%%%%%%%%%%%%%%%%%%%%

\bibliography{dipole}

\begin{thebibliography}{34}
\expandafter\ifx\csname natexlab\endcsname\relax\def\natexlab#1{#1}\fi
\expandafter\ifx\csname bibnamefont\endcsname\relax
  \def\bibnamefont#1{#1}\fi
\expandafter\ifx\csname bibfnamefont\endcsname\relax
  \def\bibfnamefont#1{#1}\fi
\expandafter\ifx\csname citenamefont\endcsname\relax
  \def\citenamefont#1{#1}\fi
\expandafter\ifx\csname url\endcsname\relax
  \def\url#1{\texttt{#1}}\fi
\expandafter\ifx\csname urlprefix\endcsname\relax\def\urlprefix{URL }\fi
\providecommand{\bibinfo}[2]{#2}
\providecommand{\eprint}[2][]{\url{#2}}

\bibitem[{\citenamefont{Seto et~al.}(2001)\citenamefont{Seto, Kawamura, and
  Nakamura}}]{bib18}
\bibinfo{author}{\bibfnamefont{N.}~\bibnamefont{Seto}},
  \bibinfo{author}{\bibfnamefont{S.}~\bibnamefont{Kawamura}}, \bibnamefont{and}
  \bibinfo{author}{\bibfnamefont{T.}~\bibnamefont{Nakamura}},
  \bibinfo{journal}{{Phys. Rev. Lett.}} \textbf{\bibinfo{volume}{{\bf{87}}}},
  \bibinfo{pages}{221103} (\bibinfo{year}{2001}).

\bibitem[{bib({\natexlab{a}})}]{bib19}
\bibinfo{note}{S. Sato {\it{et al.}}, Journal of Physics: Conference Series
  {\bf{154}}, 012040 (2009)}.

\bibitem[{bib({\natexlab{b}})}]{bib20}
\bibinfo{note}{E. S. Phinney {\it{et al.}}, The Big Bang Observer, NASA Mission
  Concept Study (2003)}.

\bibitem[{\citenamefont{Cutler and Holz}(2009)}]{bib6}
\bibinfo{author}{\bibfnamefont{C.}~\bibnamefont{Cutler}} \bibnamefont{and}
  \bibinfo{author}{\bibfnamefont{D.~E.} \bibnamefont{Holz}},
  \bibinfo{journal}{{Phys. Rev.}} \textbf{\bibinfo{volume}{{D \bf{80}}}},
  \bibinfo{pages}{104009} (\bibinfo{year}{2009}).

\bibitem[{\citenamefont{Yagi and Tanaka}(2010)}]{bib45}
\bibinfo{author}{\bibfnamefont{K.}~\bibnamefont{Yagi}} \bibnamefont{and}
  \bibinfo{author}{\bibfnamefont{T.}~\bibnamefont{Tanaka}},
  \bibinfo{journal}{{Prog. Theor. Phys.}}
  \textbf{\bibinfo{volume}{{\bf{123}}}}, \bibinfo{pages}{1069}
  (\bibinfo{year}{2010}).

\bibitem[{\citenamefont{Nishizawa et~al.}(2010)\citenamefont{Nishizawa, Taruya,
  and Kawamura}}]{bib7}
\bibinfo{author}{\bibfnamefont{A.}~\bibnamefont{Nishizawa}},
  \bibinfo{author}{\bibfnamefont{A.}~\bibnamefont{Taruya}}, \bibnamefont{and}
  \bibinfo{author}{\bibfnamefont{S.}~\bibnamefont{Kawamura}},
  \bibinfo{journal}{{Phys. Rev.}} \textbf{\bibinfo{volume}{{D \bf{81}}}},
  \bibinfo{pages}{104043} (\bibinfo{year}{2010}).

\bibitem[{\citenamefont{Schutz}(1986)}]{bib33}
\bibinfo{author}{\bibfnamefont{B.~F.} \bibnamefont{Schutz}},
  \bibinfo{journal}{{Nature (London)}} \textbf{\bibinfo{volume}{{\bf{323}}}},
  \bibinfo{pages}{310} (\bibinfo{year}{1986}).

\bibitem[{\citenamefont{Bonvin et~al.}(2006{\natexlab{a}})\citenamefont{Bonvin,
  Durrer, and Kunz}}]{bib8}
\bibinfo{author}{\bibfnamefont{C.}~\bibnamefont{Bonvin}},
  \bibinfo{author}{\bibfnamefont{R.}~\bibnamefont{Durrer}}, \bibnamefont{and}
  \bibinfo{author}{\bibfnamefont{M.}~\bibnamefont{Kunz}},
  \bibinfo{journal}{{Phys. Rev. Lett.}} \textbf{\bibinfo{volume}{{\bf{96}}}},
  \bibinfo{pages}{191302} (\bibinfo{year}{2006}{\natexlab{a}}).

\bibitem[{\citenamefont{Bonvin et~al.}(2006{\natexlab{b}})\citenamefont{Bonvin,
  Durrer, and Gasparini}}]{bib9}
\bibinfo{author}{\bibfnamefont{C.}~\bibnamefont{Bonvin}},
  \bibinfo{author}{\bibfnamefont{R.}~\bibnamefont{Durrer}}, \bibnamefont{and}
  \bibinfo{author}{\bibfnamefont{A.}~\bibnamefont{Gasparini}},
  \bibinfo{journal}{{Phys. Rev.}} \textbf{\bibinfo{volume}{{D \bf{73}}}},
  \bibinfo{pages}{023523} (\bibinfo{year}{2006}{\natexlab{b}}).

\bibitem[{\citenamefont{Stern et~al.}(2010)\citenamefont{Stern, Jimenez, Verde,
  Kamionkowski, and Stanford}}]{bib24}
\bibinfo{author}{\bibfnamefont{D.}~\bibnamefont{Stern}},
  \bibinfo{author}{\bibfnamefont{R.}~\bibnamefont{Jimenez}},
  \bibinfo{author}{\bibfnamefont{L.}~\bibnamefont{Verde}},
  \bibinfo{author}{\bibfnamefont{M.}~\bibnamefont{Kamionkowski}},
  \bibnamefont{and} \bibinfo{author}{\bibfnamefont{S.~A.}
  \bibnamefont{Stanford}}, \bibinfo{journal}{{J. Cosmol. Astropart. Phys.}}
  \textbf{\bibinfo{volume}{{\bf{02}}}}, \bibinfo{pages}{008}
  (\bibinfo{year}{2010}).

\bibitem[{\citenamefont{Gaztanaga
  et~al.}(2009{\natexlab{a}})\citenamefont{Gaztanaga, Miquel, and
  Sanchez}}]{bib32}
\bibinfo{author}{\bibfnamefont{E.}~\bibnamefont{Gaztanaga}},
  \bibinfo{author}{\bibfnamefont{R.}~\bibnamefont{Miquel}}, \bibnamefont{and}
  \bibinfo{author}{\bibfnamefont{E.}~\bibnamefont{Sanchez}},
  \bibinfo{journal}{{Phys. Rev. Lett.}} \textbf{\bibinfo{volume}{{\bf{103}}}},
  \bibinfo{pages}{091302} (\bibinfo{year}{2009}{\natexlab{a}}).

\bibitem[{\citenamefont{Gaztanaga
  et~al.}(2009{\natexlab{b}})\citenamefont{Gaztanaga, Cabre, and Hui}}]{bib40}
\bibinfo{author}{\bibfnamefont{E.}~\bibnamefont{Gaztanaga}},
  \bibinfo{author}{\bibfnamefont{A.}~\bibnamefont{Cabre}}, \bibnamefont{and}
  \bibinfo{author}{\bibfnamefont{L.}~\bibnamefont{Hui}},
  \bibinfo{journal}{{Mon. Not. R. Astron. Soc.}}
  \textbf{\bibinfo{volume}{{\bf{399}}}}, \bibinfo{pages}{1663}
  (\bibinfo{year}{2009}{\natexlab{b}}).

\bibitem[{\citenamefont{Sasaki}(1987)}]{bib55}
\bibinfo{author}{\bibfnamefont{M.}~\bibnamefont{Sasaki}},
  \bibinfo{journal}{{Mon. Not. R. Astron. Soc.}}
  \textbf{\bibinfo{volume}{{\bf{228}}}}, \bibinfo{pages}{653}
  (\bibinfo{year}{1987}).

\bibitem[{bib({\natexlab{c}})}]{bib13}
\bibinfo{note}{N. Jarosik et al., arXiv:1001.4744.}

\bibitem[{\citenamefont{Corasaniti et~al.}(2006)\citenamefont{Corasaniti,
  Verde, Crotts, and Blake}}]{bib53}
\bibinfo{author}{\bibfnamefont{P.~S.} \bibnamefont{Corasaniti}},
  \bibinfo{author}{\bibfnamefont{M.~L.} \bibnamefont{Verde}},
  \bibinfo{author}{\bibfnamefont{A.}~\bibnamefont{Crotts}}, \bibnamefont{and}
  \bibinfo{author}{\bibfnamefont{C.}~\bibnamefont{Blake}},
  \bibinfo{journal}{{Mon. Not. R. Astron. Soc.}}
  \textbf{\bibinfo{volume}{{\bf{369}}}}, \bibinfo{pages}{798}
  (\bibinfo{year}{2006}).

\bibitem[{bib({\natexlab{d}})}]{bib54}
\bibinfo{note}{A. Crotts {\it{et al.}}, arXiv:astro-ph/0507043.}

\bibitem[{\citenamefont{Cutler and Flanagan}(1994)}]{bib46}
\bibinfo{author}{\bibfnamefont{C.}~\bibnamefont{Cutler}} \bibnamefont{and}
  \bibinfo{author}{\bibfnamefont{E.~E.} \bibnamefont{Flanagan}},
  \bibinfo{journal}{{Phys. Rev.}} \textbf{\bibinfo{volume}{{D \bf{49}}}},
  \bibinfo{pages}{2658} (\bibinfo{year}{1994}).

\bibitem[{bib({\natexlab{e}})}]{bib47}
\bibinfo{note}{M. Maggiore, {\it{Gravitational Waves}} (Oxford university
  press, 2008)}.

\bibitem[{\citenamefont{Takahashi and Nakamura}(2005)}]{bib5}
\bibinfo{author}{\bibfnamefont{R.}~\bibnamefont{Takahashi}} \bibnamefont{and}
  \bibinfo{author}{\bibfnamefont{T.}~\bibnamefont{Nakamura}},
  \bibinfo{journal}{{Prog. Theor. Phys.}}
  \textbf{\bibinfo{volume}{{\bf{113}}}}, \bibinfo{pages}{63}
  (\bibinfo{year}{2005}).

\bibitem[{\citenamefont{Finn}(1992)}]{bib48}
\bibinfo{author}{\bibfnamefont{L.~S.} \bibnamefont{Finn}},
  \bibinfo{journal}{{Phys. Rev.}} \textbf{\bibinfo{volume}{{D \bf{46}}}},
  \bibinfo{pages}{5236} (\bibinfo{year}{1992}).

\bibitem[{\citenamefont{Prince et~al.}(2002)\citenamefont{Prince, Tinto,
  Larson, and Armstrong}}]{bib21}
\bibinfo{author}{\bibfnamefont{T.~A.} \bibnamefont{Prince}},
  \bibinfo{author}{\bibfnamefont{M.}~\bibnamefont{Tinto}},
  \bibinfo{author}{\bibfnamefont{S.~L.} \bibnamefont{Larson}},
  \bibnamefont{and} \bibinfo{author}{\bibfnamefont{J.~W.}
  \bibnamefont{Armstrong}}, \bibinfo{journal}{{Phys. Rev.}}
  \textbf{\bibinfo{volume}{D {\bf{66}}}}, \bibinfo{pages}{122002}
  (\bibinfo{year}{2002}).

\bibitem[{\citenamefont{Cutler and Harms}(2006)}]{bib11}
\bibinfo{author}{\bibfnamefont{C.}~\bibnamefont{Cutler}} \bibnamefont{and}
  \bibinfo{author}{\bibfnamefont{J.}~\bibnamefont{Harms}},
  \bibinfo{journal}{{Phys. Rev.}} \textbf{\bibinfo{volume}{{D \bf{73}}}},
  \bibinfo{pages}{042001} (\bibinfo{year}{2006}).

\bibitem[{\citenamefont{Schneider et~al.}(2001)\citenamefont{Schneider,
  Ferrari, Matarrese, and Zwart}}]{bib10}
\bibinfo{author}{\bibfnamefont{R.}~\bibnamefont{Schneider}},
  \bibinfo{author}{\bibfnamefont{V.}~\bibnamefont{Ferrari}},
  \bibinfo{author}{\bibfnamefont{S.}~\bibnamefont{Matarrese}},
  \bibnamefont{and} \bibinfo{author}{\bibfnamefont{S.~F.~P.}
  \bibnamefont{Zwart}}, \bibinfo{journal}{{Mon. Not. R. Astron. Soc.}}
  \textbf{\bibinfo{volume}{{\bf{324}}}}, \bibinfo{pages}{797}
  (\bibinfo{year}{2001}).

\bibitem[{\citenamefont{Abadie et~al.}(2010)}]{bib12}
\bibinfo{author}{\bibfnamefont{J.}~\bibnamefont{Abadie}} \bibnamefont{et~al.},
  \bibinfo{journal}{{Classical Quantum Gravity}}
  \textbf{\bibinfo{volume}{{27}}}, \bibinfo{pages}{173001}
  (\bibinfo{year}{2010}).

\bibitem[{\citenamefont{Wambsganss et~al.}(1997)\citenamefont{Wambsganss, Cen,
  Xu, and Ostriker}}]{bib51}
\bibinfo{author}{\bibfnamefont{J.}~\bibnamefont{Wambsganss}},
  \bibinfo{author}{\bibfnamefont{R.}~\bibnamefont{Cen}},
  \bibinfo{author}{\bibfnamefont{G.}~\bibnamefont{Xu}}, \bibnamefont{and}
  \bibinfo{author}{\bibfnamefont{J.~P.} \bibnamefont{Ostriker}},
  \bibinfo{journal}{{Astrophys. J.}} \textbf{\bibinfo{volume}{{\bf{475}}}},
  \bibinfo{pages}{L81} (\bibinfo{year}{1997}).

\bibitem[{\citenamefont{Holz and Linder}(2005)}]{bib14}
\bibinfo{author}{\bibfnamefont{D.~E.} \bibnamefont{Holz}} \bibnamefont{and}
  \bibinfo{author}{\bibfnamefont{E.~V.} \bibnamefont{Linder}},
  \bibinfo{journal}{Astrophys. J.} \textbf{\bibinfo{volume}{{\bf{631}}}},
  \bibinfo{pages}{678} (\bibinfo{year}{2005}).

\bibitem[{\citenamefont{Holz and Wald}(1998)}]{bib50}
\bibinfo{author}{\bibfnamefont{D.~E.} \bibnamefont{Holz}} \bibnamefont{and}
  \bibinfo{author}{\bibfnamefont{R.~M.} \bibnamefont{Wald}},
  \bibinfo{journal}{{Phys. Rev.}} \textbf{\bibinfo{volume}{{D \bf{58}}}},
  \bibinfo{pages}{063501} (\bibinfo{year}{1998}).

\bibitem[{bib({\natexlab{f}})}]{bib49}
\bibinfo{note}{K. Kainulainen and V. Marra, arXiv:1011.0732.}

\bibitem[{\citenamefont{Hirata et~al.}(2010)\citenamefont{Hirata, Holz, and
  Cutler}}]{bib15}
\bibinfo{author}{\bibfnamefont{C.~M.} \bibnamefont{Hirata}},
  \bibinfo{author}{\bibfnamefont{D.~E.} \bibnamefont{Holz}}, \bibnamefont{and}
  \bibinfo{author}{\bibfnamefont{C.}~\bibnamefont{Cutler}},
  \bibinfo{journal}{{Phys. Rev.}} \textbf{\bibinfo{volume}{D {\bf{81}}}},
  \bibinfo{pages}{124046} (\bibinfo{year}{2010}).

\bibitem[{\citenamefont{Shang and Haiman}(2011)}]{bib16}
\bibinfo{author}{\bibfnamefont{C.}~\bibnamefont{Shang}} \bibnamefont{and}
  \bibinfo{author}{\bibfnamefont{Z.}~\bibnamefont{Haiman}},
  \bibinfo{journal}{{Mon. Not. R. Astron. Soc.}}
  \textbf{\bibinfo{volume}{{\bf{411}}}}, \bibinfo{pages}{9}
  (\bibinfo{year}{2011}).

\bibitem[{\citenamefont{Gordon et~al.}(2007)\citenamefont{Gordon, Land, and
  Slosar}}]{bib17}
\bibinfo{author}{\bibfnamefont{C.}~\bibnamefont{Gordon}},
  \bibinfo{author}{\bibfnamefont{K.}~\bibnamefont{Land}}, \bibnamefont{and}
  \bibinfo{author}{\bibfnamefont{A.}~\bibnamefont{Slosar}},
  \bibinfo{journal}{{Phys. Rev. Lett.}} \textbf{\bibinfo{volume}{{\bf{99}}}},
  \bibinfo{pages}{081301} (\bibinfo{year}{2007}).

\bibitem[{\citenamefont{Silberman et~al.}(2001)\citenamefont{Silberman, Dekel,
  Eldar, and Zehavi}}]{bib52}
\bibinfo{author}{\bibfnamefont{L.}~\bibnamefont{Silberman}},
  \bibinfo{author}{\bibfnamefont{A.}~\bibnamefont{Dekel}},
  \bibinfo{author}{\bibfnamefont{A.}~\bibnamefont{Eldar}}, \bibnamefont{and}
  \bibinfo{author}{\bibfnamefont{I.}~\bibnamefont{Zehavi}},
  \bibinfo{journal}{{Astrophys. J.}} \textbf{\bibinfo{volume}{{\bf{557}}}},
  \bibinfo{pages}{102} (\bibinfo{year}{2001}).

\bibitem[{\citenamefont{Shapiro et~al.}(2010)\citenamefont{Shapiro, Bacon,
  Hendry, and Hoyle}}]{bib43}
\bibinfo{author}{\bibfnamefont{C.}~\bibnamefont{Shapiro}},
  \bibinfo{author}{\bibfnamefont{D.~J.} \bibnamefont{Bacon}},
  \bibinfo{author}{\bibfnamefont{M.}~\bibnamefont{Hendry}}, \bibnamefont{and}
  \bibinfo{author}{\bibfnamefont{B.}~\bibnamefont{Hoyle}},
  \bibinfo{journal}{{Mon. Not. R. Astron. Soc.}}
  \textbf{\bibinfo{volume}{{\bf{404}}}}, \bibinfo{pages}{858}
  (\bibinfo{year}{2010}).

\bibitem[{\citenamefont{Hilbert et~al.}(2011)\citenamefont{Hilbert, Gair, and
  King}}]{bib44}
\bibinfo{author}{\bibfnamefont{S.}~\bibnamefont{Hilbert}},
  \bibinfo{author}{\bibfnamefont{J.~R.} \bibnamefont{Gair}}, \bibnamefont{and}
  \bibinfo{author}{\bibfnamefont{L.~J.} \bibnamefont{King}},
  \bibinfo{journal}{{Mon. Not. R. Astron. Soc.}}
  \textbf{\bibinfo{volume}{{\bf{412}}}}, \bibinfo{pages}{1023}
  (\bibinfo{year}{2011}).

\end{thebibliography}

\end{document}